\documentclass[nofootinbib,reprint,floatfix, longbibliography, superscriptaddress]{revtex4-1}
\usepackage{amsmath}
\usepackage{amssymb}
\usepackage{graphicx}
\usepackage{verbatim}
\usepackage{color}
\usepackage{subfigure}
\usepackage[pdftex, breaklinks,colorlinks, citecolor=blue, urlcolor=blue]{hyperref}

\begin{document}

\title{Quantum-enabled temporal and spectral mode conversion of microwave signals}
\author{R. W. Andrews}
\affiliation{JILA, University of Colorado and NIST, Boulder, Colorado 80309, USA}
\affiliation{Department of Physics, University of Colorado, Boulder, Colorado 80309, USA}

\author{A. P. Reed}
\affiliation{JILA, University of Colorado and NIST, Boulder, Colorado 80309, USA}
\affiliation{Department of Physics, University of Colorado, Boulder, Colorado 80309, USA}

\author{K. Cicak}
\affiliation{National Institute of Standards and Technology (NIST), Boulder, Colorado 80305, USA}

\author{J. D. Teufel}
\affiliation{National Institute of Standards and Technology (NIST), Boulder, Colorado 80305, USA}

\author{K. W. Lehnert}
\affiliation{JILA, University of Colorado and NIST, Boulder, Colorado 80309, USA}
\affiliation{Department of Physics, University of Colorado, Boulder, Colorado 80309, USA}
\affiliation{National Institute of Standards and Technology (NIST), Boulder, Colorado 80305, USA}

\maketitle

Electromagnetic waves are ideal candidates for transmitting information in a quantum network \cite{jelena2009} as they can be routed rapidly and efficiently between locations using optical fibers or microwave cables.  Yet linking quantum-enabled devices with cables has proved difficult because most cavity or circuit quantum electrodynamics (cQED) systems used in quantum information processing can only absorb and emit signals with a specific frequency and temporal envelope \cite{houck2007, santori2002}.  Here we show that the temporal and spectral content of microwave-frequency electromagnetic signals can be arbitrarily manipulated with a flexible aluminum drumhead embedded in a microwave circuit.  The aluminum drumhead simultaneously forms a mechanical oscillator and a tunable capacitor. This device offers a way to build quantum microwave networks using separate and otherwise mismatched components.  Furthermore, it will enable the preparation of non-classical states of motion by capturing non-classical microwave signals prepared by the most coherent circuit QED systems.

Cavity quantum electrodynamics and its low-frequency on-chip counterpart circuit quantum electrodynamics form prototypical quantum systems that have enabled many tests of quantum theory and advanced goals of quantum computation \cite{kimble2005,girvin2008}.  A single cQED system links a quantum two-level system (qubit) to excitations of the electromagnetic field. Linking multiple, spatially separated cQED systems with electromagnetic waves provides a scalable quantum networking architecture that could combine a large number of individual systems (or nodes) into one network \cite{kimble2008,rempe2012}.  Deterministically distributing information between different nodes requires the electromagnetic signals generated by one node be completely absorbed by another.  This requirement can be viewed as a generalized impedance matching, which can be satisfied by controlling the temporal and spectral content of the signal \cite{mabuchi1997, korotkov2011}.  

The radiation pressure interaction between light and a vibrating mass provides a natural way to manipulate electromagnetic signals.  It has been used to store electromagnetic signals \cite{verhagen2012,palomaki2013}, amplify them \cite{massel2011}, and convert them between different frequencies \cite{hill2012, andrews2014}.  By creating a tunable circuit that utilizes the radiation pressure interaction, we combine microwave frequency pulse shaping \cite{yin2013,houck2014,pechal2014} and conversion \cite{abdo2013, aumentado2011, inomata2014} between adjustable frequencies.  As a demonstration of our temporal and spectral mode converter, we use it to implement a particular protocol that converts the temporal envelope of a 7 GHz microwave signal from a decaying exponential to a Gaussian, and also shifts the signal's frequency by up to 250~MHz.  Such a protocol can process signals emitted from coherent circuit QED systems, which emit signals in a fixed, narrow ($\lesssim 10$~MHz) frequency range, and with an exponentially decaying envelope \cite{houck2007}.  To test the suitability of the mode converter for use with quantum signals, we characterize the noise acquired by low-amplitude classical signals during temporal and spectral mode conversion.  We find the total added noise is less than one quantum for frequency conversions up to approximately $100$ MHz, suggesting that mode conversion is suitable for use with fragile quantum signals \cite{ulf1997}.

\begin{figure}
\begin{minipage}{\linewidth}
\scalebox{.5}{\includegraphics{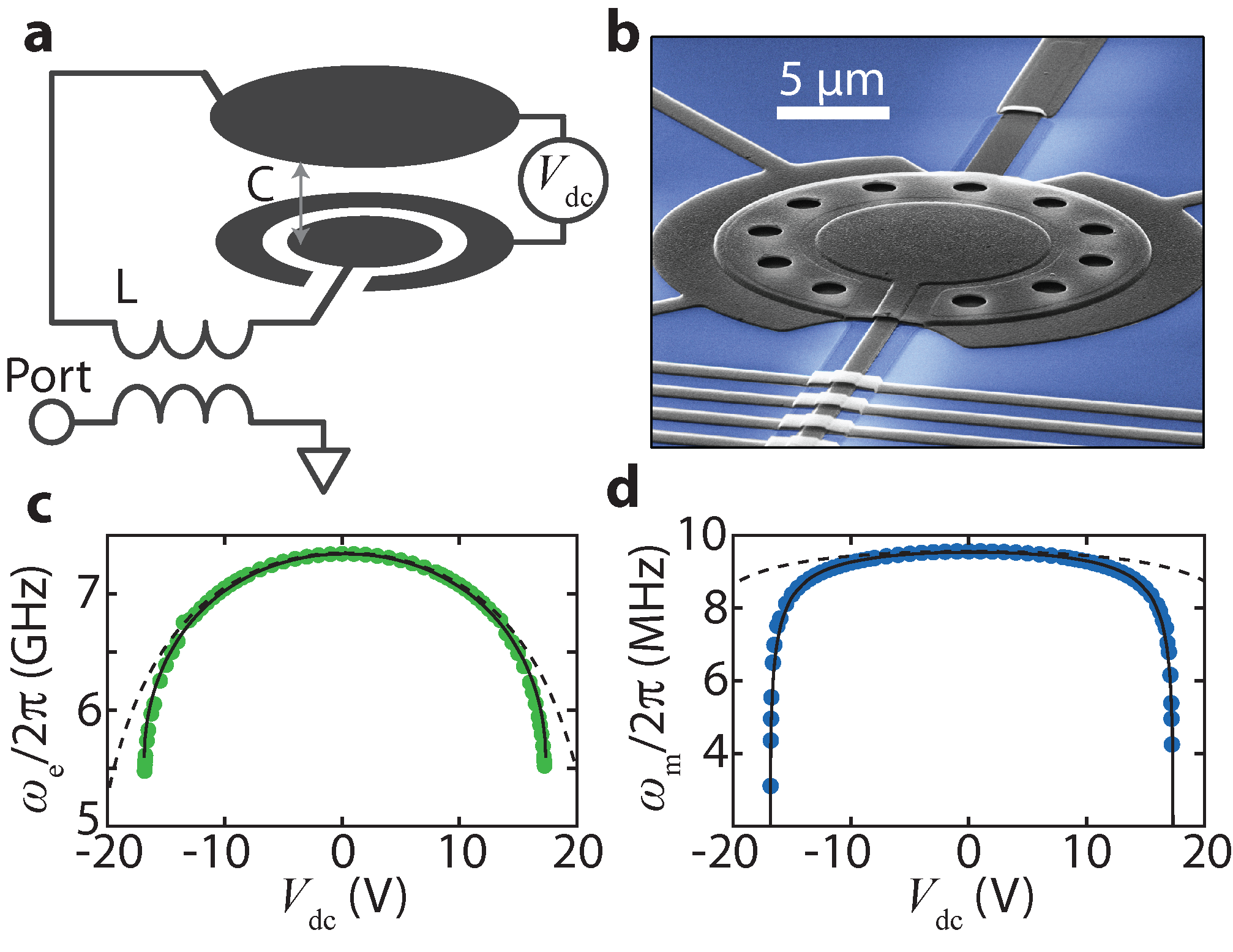}}
\caption{A tunable electromechanical circuit.  {\bf{a}}, An inductor (L) and a parallel plate capacitor (C) form a microwave-frequency electromagnetic resonator that is inductively coupled to a transmission line (Port).  The upper plate of the parallel-plate capacitor is clamped only on the edges, and is free to vibrate like a drumhead.  The capacitance of the microwave circuit can be tuned by applying a voltage ($V_{\mathrm{dc}}$) between an annular electrode and the drumhead.  {\bf{b}}, A scanning electron micrograph of the parallel plate capacitor, formed by sputtering layers of aluminum (dark gray) onto a sapphire substrate (blue); the micrograph is oriented similarly to panel {\bf{a}}. {\bf{c}}, {\bf{d}}, Measured microwave resonant frequency ($\omega_{\mathrm{e}}$) and mechanical resonant frequency ($\omega_{\mathrm{m}}$) as a function of $V_{\mathrm{dc}}$.  Solid lines show the expected performance as calculated with a finite element simulation that includes an attractive force that scales like the Casimir force and is within a factor of two of an estimated Casimir force magnitude \cite{intravaia2005}; neglecting this additional attractive force results in much different behavior shown with the dashed lines.  Although incorporation of the Casimir force accounts for our data, we can not rule out some contribution from patch potentials \cite{speake2003}.}
\label{devicev1}
\end{minipage}
\end{figure}

The mode converter relies on a mechanical oscillator formed by a flexible aluminum drumhead, as shown in Fig.~\ref{devicev1} \cite{cicak2010}.  The drumhead is an integral part of a microwave circuit that is addressed with a coaxial transmission line.  During operation, the circuit is mounted on the base of a dilution refrigerator, and cooled to $<$25 millikelvin.    
Applying a voltage, $V_{\mathrm{dc}}$, between the drumhead and an additional annular electrode (Fig.~\ref{devicev1}a) creates an attractive force between the two metal films.  As the drumhead is free to move, it deflects and increases the total capacitance of the microwave circuit, and thus lowers its resonant frequency, $\omega_{\mathrm{e}}$, as shown in Fig.~\ref{devicev1}c.  The separation between the drumhead and microwave electrode is approximately $40$ nm with $V_{\mathrm{dc}}=0$, and at this separation each nanometer of drumhead motion alters the resonant frequency of the microwave circuit by approximately 50 MHz, or an electromechanical coupling of $G\approx2\pi\times 50$~ MHz/nm.  By using the electrostatic force to move the drumhead approximately 20 nm, we shift the circuit's resonant frequency by over 1 GHz, much more than the circuit's bandwidth of $\kappa=2\pi\times2.5$~MHz.  As $V_{\mathrm{dc}}$ is increased, the electrostatic force and other attractive forces eventually overwhelm the restoring force provided by tension in the drumhead, evidenced by the decrease in the vibrational frequency, $\omega_{\mathrm{m}}$, as shown in Fig.~\ref{devicev1}d.  When the vibrational frequency reaches zero, the attractive forces dominate and the drumhead will collapse.  

Vibrations of the drumhead are also affected by the electromechanical coupling $G$.   At $V_{\mathrm{dc}}=0$, $G=2\pi\times 42$~MHz/nm, and in the presence of a detuned microwave pump, this coupling exchanges excitations of the microwave circuit and vibrational excitations of the drumhead at a rate $g(t)=G x_{\mathrm{zp}}\sqrt{n(t)}$  \cite{palomaki2013}, where $x_{\mathrm{zp}}=6.4 ~\mathrm{fm}$ is the zero-point motion of the mechanical oscillator and $n(t)$ is the strength of the microwave pump expressed as the number of photons induced in the circuit by the pump.  We choose to work in the weak-coupling regime, where excitations in the microwave circuit decay into the transmission line (see Fig.~\ref{devicev1}a) at a rate $\kappa_{\mathrm{ext}}> 2g$.  In this regime, the microwave circuit simply enhances an exchange between excitations in the transmission line and vibrational excitations of the drumhead \cite{zhang2003}; this exchange occurs at a rate $\Gamma(t)=4g^2/\kappa$, where $\kappa$ is the total energy decay rate of the microwave circuit.  The controllable, time dependent coupling to a transmission line provided by this device  is a basic requirement for controlling the temporal envelope of signals \cite{korotkov2011}.

\begin{figure}
\begin{minipage}{\linewidth}
\scalebox{.8333}{\includegraphics{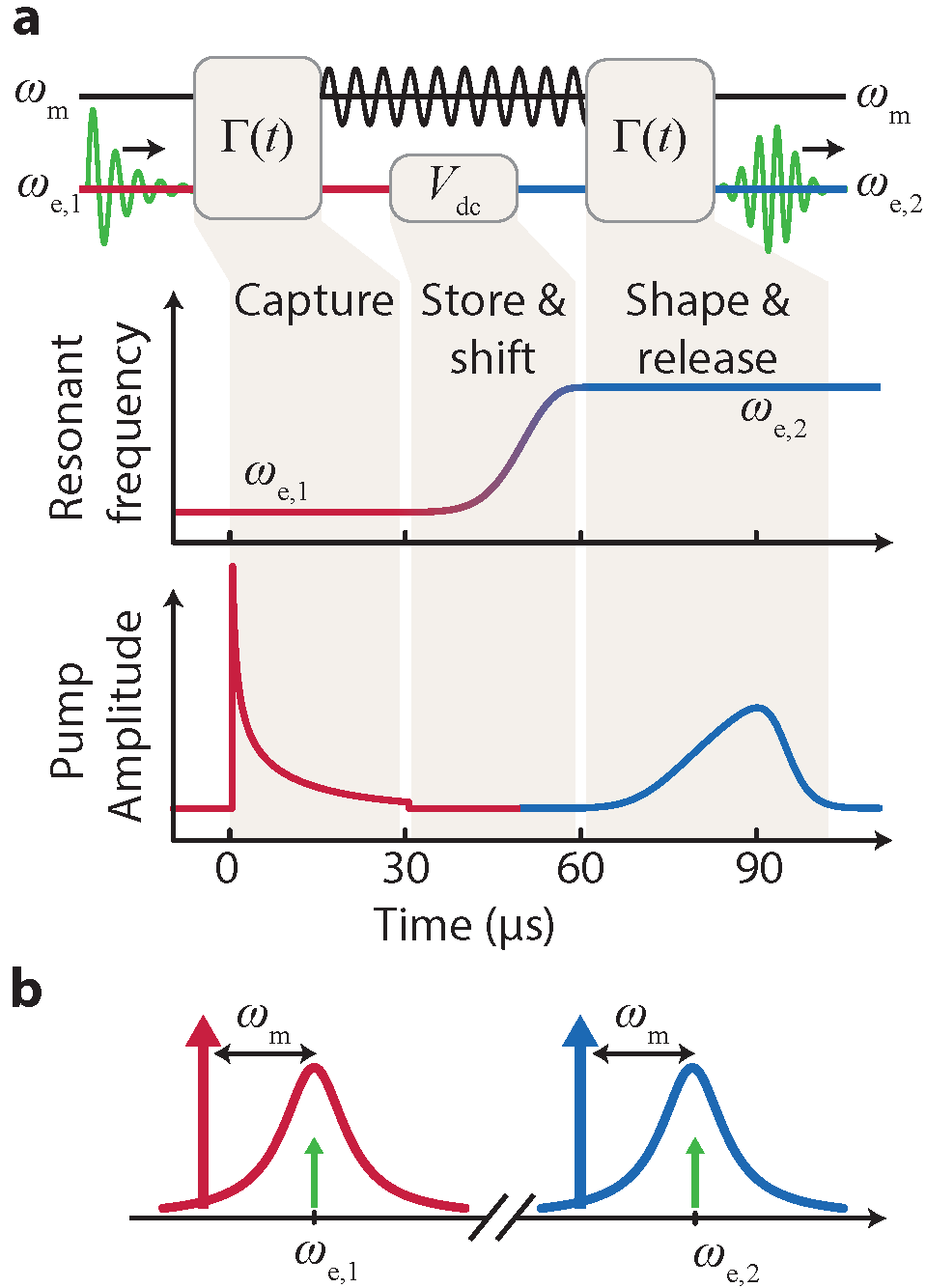}}
\caption{Protocol for mode conversion.  {\bf{a}}, A signal in the transmission line at frequency $\omega_{\mathrm{e,1}}$ (green sinusoid) is first converted into a vibration of the aluminum drumhead (black sinusoid).  $\Gamma(t)$ is adjusted by varying the power in a pump at frequency  $\omega_{\mathrm{e,1}}-\omega_{\mathrm{m}}$ (red waveform).  Then, we alter the frequency of the microwave circuit by changing $V_{\mathrm{dc}}$.  During this time, the signal is stored in the vibrating drumhead; the change in $V_{\mathrm{dc}}$ adiabatically changes $\omega_{\mathrm{m}}$ by approximately $2\pi\times300$ kHz.  Finally, the signal is transferred from the drumhead back into the transmission line at frequency $\omega_{\mathrm{e,2}}$ by adjusting the power in a pump at frequency $\omega_{\mathrm{e,2}}-\omega_{\mathrm{m}}$ (blue waveform). {\bf{b}}, Frequency domain schematic of  mode conversion.  Microwave pumps (red and blue arrows) applied at a frequency $\omega_{\mathrm{m}}$ below the microwave circuit (response shown as the red and blue curves) allow a signal at frequency $\omega_{\mathrm{e,1}}$ to be converted to a signal at frequency $\omega_{\mathrm{e,2}}$ (green arrows).   }
\label{protocolv1}
\end{minipage}
\end{figure}

Our protocol for mode conversion consists of modulating both $V_{\mathrm{dc}}$ and $\Gamma (t)$.  Initially, we modulate $\Gamma (t)$ to capture a microwave signal propagating down the transmission line and store it as a vibration of the drumhead \cite{harlow2013}.  The signal can be stored as a vibration for over 100 $\mu$s without being lost or corrupted, which provides sufficient time to complete the mode conversion protocol, and also serves as an adjustable signal storage or delay element.  Once the signal is captured, we change $V_{\mathrm{dc}}$ to alter the microwave circuit's resonant frequency. We then modulate $\Gamma (t)$ to release the signal stored in the drumhead back into the transmission line.  Even though the released signal has different spectral and temporal content from the initial signal, the information in the initial signal has ideally been preserved.  The protocol is schematically depicted in Fig.~\ref{protocolv1}.

\begin{figure*}
\begin{minipage}{\linewidth}
\scalebox{.33}{\includegraphics{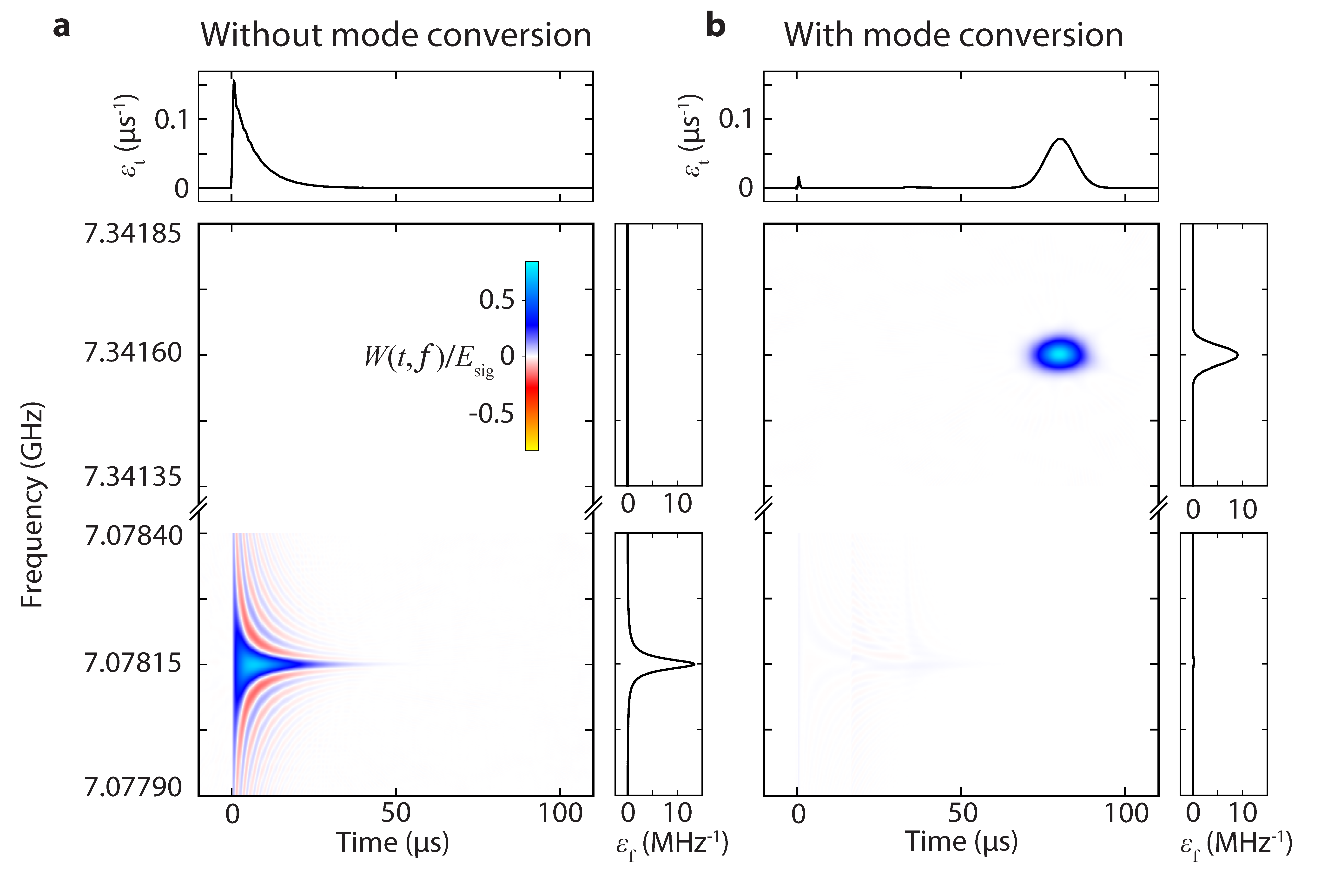}}
\caption{Temporal and spectral mode conversion of a microwave signal.  {\bf{a}}, Wigner-Ville distribution of a signal that reflects off the mode converter (so no mode conversion is performed) and is measured with a microwave receiver.  The marginal distributions of the Wigner-Ville distribution show the signal is a decaying exponential in time, and a Lorentzian in frequency;  $\epsilon_{\mathrm{t}}=|v(t)|^2/E_{\mathrm{sig}}$ and $\epsilon_{\mathrm{f}}=|v(f)|^2/E_{\mathrm{sig}}$.  The signal is centered near 7.07815 GHz and has a power decay rate of $2\pi\times 24 ~\mathrm{kHz}$. There is no signal content near 7.34160 GHz.  {\bf{b}}, Wigner-Ville distribution of a mode-converted signal.  When the mode converter is used, the decaying exponential signal is captured as a vibration of the drumhead, and the signal content at 7.07815 GHz is almost entirely absent.  Later in time, we recover the signal that is now centered near 7.34160 GHz and has a Gaussian temporal envelope and Gaussian spectral content with width $2\pi\times24$ kHz.  }
\label{wigville1}
\end{minipage}
\end{figure*}

Signals reflected or emitted from the mode converter are measured with a microwave receiver that consists of a Josephson parametric amplifier \cite{manuel2008} followed by additional amplifiers, a down-converting mixer, and a digitizer (see S.I.).  Collected data consist of time-stamped voltage values that we coherently average by running the protocol multiple times.  To clearly represent the time and frequency content of a measured voltage waveform, we construct the Wigner-Ville distribution using the discretized version of $W(t, f)=\int_{-\infty}^{\infty}v(t+\tau/2)v^{*}(t-\tau/2)e^{ 2\pi i f \tau}d\tau$, where $W(t, f)$ is the Wigner-Ville distribution and $v(t)$ is the analytic voltage measured at time $t$.  The analytic voltage is constructed using a Hilbert transform on the measured voltage \cite{boashash1988}: $v(t)=V(t)+i \mathcal{H}(V(t))$, where $V(t)$ is the voltage measured at time $t$ and $\mathcal{H}$ is the Hilbert transform.  The resulting time-frequency distribution is similar to the position-momentum Wigner quasiprobability distribution; however, nothing `quantum' is associated with a negative distribution.  The marginal distributions of the Wigner-Ville distribution give the temporal envelope and energy spectral density of the voltage waveform.

As an initial test of our measurement apparatus, we generate a signal near 7.08 GHz that has an exponentially decaying envelope (power decay rate of $2\pi\times 24$ kHz) and inject it into the coaxial transmission line that connects to the mode converter.  We choose a signal with this temporal envelope because it resembles the temporal envelope of quantum signals emitted from highly coherent circuit QED systems \cite{houck2007}.  We set $V_{\mathrm{dc}}$ so that the microwave circuit has a resonant frequency substantially different from the injected signal, and the signal reflects off the mode converter and is measured by our microwave receiver.  The measured Wigner-Ville distribution of the injected signal is shown in Fig.~\ref{wigville1}a, with the data normalized by $E_{\mathrm{sig}}=\int_{-\infty}^\infty |v_{\mathrm{sig}}(t)|^2 dt$, the total energy in the unconverted signal.

To perform mode conversion, we set $V_{\mathrm{dc}}\approx10~ \mathrm{V}$ so the microwave circuit is centered near 7.08 GHz, and send in the signal with the decaying exponential envelope.  We apply a microwave pump with a time-dependent amplitude as shown in Fig.~\ref{protocolv1}a to alter $\Gamma(t)$.  The pump amplitude is carefully chosen \cite{harlow2013} so that we capture almost all of the signal and store it as a vibration of the drumhead.  Only a small amount (about 4\%) of the signal energy is reflected out of the converter, as demonstrated in Fig.~\ref{wigville1}b.  We then set $V_{\mathrm{dc}}=0$ so that the microwave circuit is centered near 7.34 GHz, and again apply a microwave pump with a time dependent amplitude as shown in Fig.~\ref{protocolv1}a.  The Wigner-Ville distribution of the signal emitted by the mode converter is plotted in Fig.~\ref{wigville1}b.  From the relative energies in the reflected and converted signals, we find the photon number efficiency of the conversion process to be 0.81, in good agreement with an expected efficiency of $\left( \kappa_{\mathrm{ext}}/\kappa\right)^2\times 0.96=0.81$ where 0.96 is the efficiency with which we capture the injected signal \cite{harlow2013}.

The mode converter can be used to change signals between arbitrary temporal and spectral modes.  We demonstrated frequency conversion over approximately $250~ \mathrm{MHz}$; conversion over a larger window is feasible but the mode converter becomes more sensitive to fluctuations in $V_{\mathrm{dc}}$ at higher bias.  Although we altered the temporal envelope of a microwave signal from a decaying exponential to a Gaussian, other envelopes are possible, only limited by the bandwidth of the mode converter.  The maximum bandwidth is set by the rate at which information can be exchanged between the transmission line and the drumhead, which for this device is limited by the microwave circuit's bandwidth of $\kappa/2 \approx 2\pi\times 1~ \mathrm{MHz}$.  The minimum bandwidth is set by the mechanical decoherence rate.  Even though energy in the drumhead is lost to the environment at the modest rate of $\kappa_\mathrm{m}=2\pi\times 25$ Hz, fluctuations of the environment incoherently drive vibrations of the drumhead.  To avoid these environment-driven vibrations, signals must be processed faster than the mechanical decoherence rate of $n_{\mathrm{m}}\kappa_\mathrm{m} =2\pi \times900$~Hz, where $n_{\mathrm{m}}$ is the thermal equilibrium phonon occupation number of the drumhead (S.I.).  Avoiding these excess vibrations is essential when manipulating fragile quantum signals.  

\begin{figure}
\begin{minipage}{\linewidth}
\scalebox{.8333}{\includegraphics{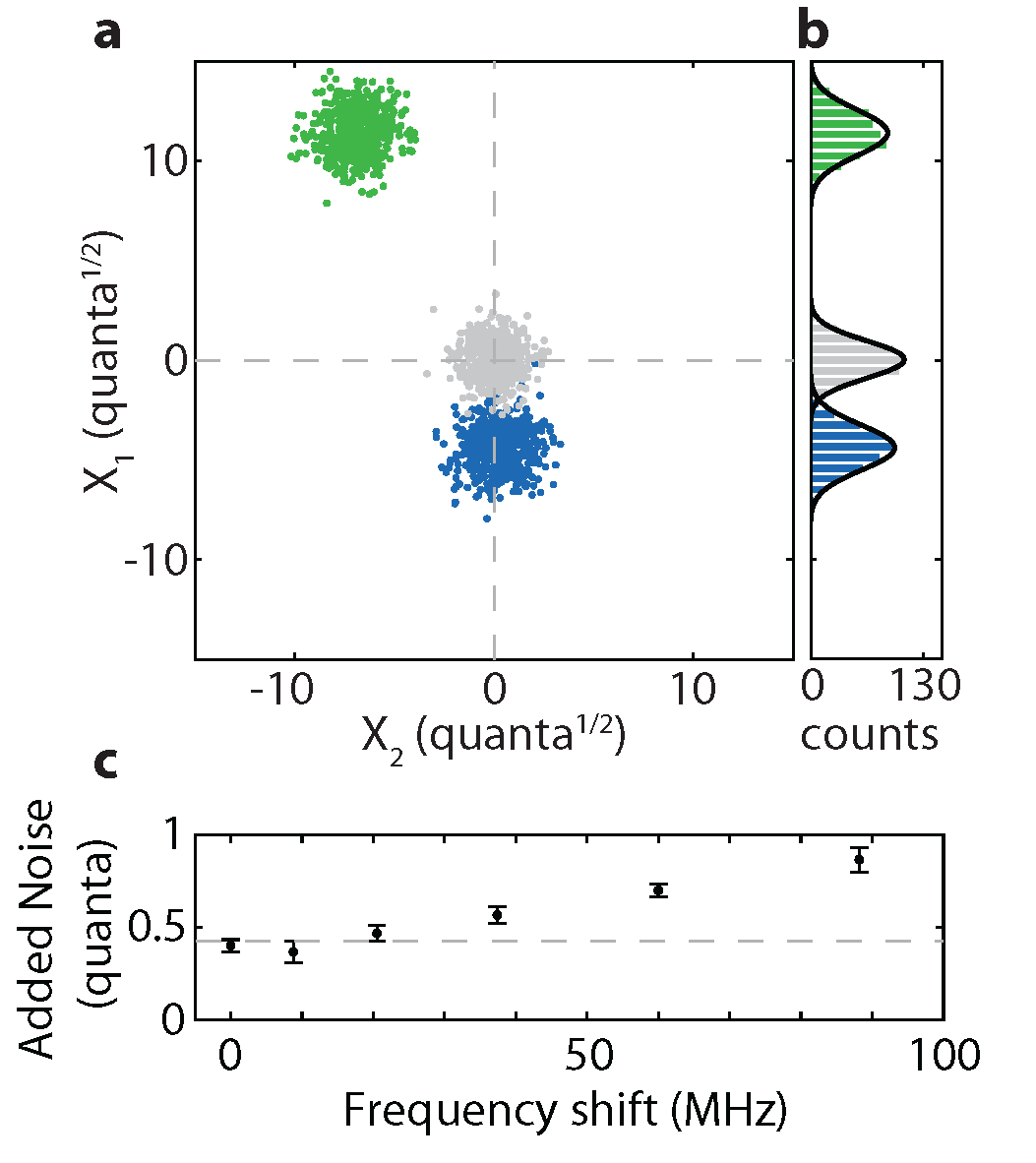}}
\caption{Noise added during mode conversion.  {\bf{a}}, Quadrature amplitudes for 500 independent repetitions of the mode conversion protocol, found using least-squares parameter estimation (S.I.).  Results of conversion with a 90 MHz frequency shift with a signal containing $\sim$$10~ \mathrm{quanta}$ (blue points) and $\sim$$100~ \mathrm{quanta}$ (green points) of energy, which demonstrates the dynamic range of the converter.  Low and high amplitude signals are intentionally generated with different phases.  For reference, a vacuum input into our microwave receiver yields the gray points.  The total variance of the quadrature amplitudes, Var(X$_1$)+Var(X$_2$), subtracted from the variance of measured vacuum indicates the amount of noise added during signal preparation and mode conversion, which we find to be $0.92 \pm 0.07$ quanta. {\bf{b}}, $X_{1}$ marginal of the quadrature amplitudes with Gaussian fits (black curves).  {\bf{c}}, For frequency shifts less than 90 MHz, we observe less added noise. Error bars, s.e.m.  Dashed gray line indicates expected added noise from a mechanical decoherence rate of $n_{\mathrm{m}}\kappa_\mathrm{m}=2\pi\times900$ Hz.  }
\label{noise}
\end{minipage}
\end{figure}

To verify that our converter operates quickly and quietly enough to avoid corrupting signals, we collect statistics on single-shot measurements of mode conversion using low amplitude ($\sim$10 quanta of energy) signals.  We inject a low-amplitude signal  with a decaying exponential envelope into the converter, shift its frequency by 90 MHz, and recover a signal with a Gaussian envelope.  The central frequency and temporal envelope of the converted signal are already known, and we use this information to perform a least-squares parameter estimation of the quadrature amplitudes X$_{\mathrm{1}}$ and X$_{2}$ (S.I.).  The quadrature amplitudes are two independent values that can be recast as the amplitude and phase of the converted signal.  The result of each single shot measurement and quadrature amplitude estimation appears as a single point in Fig.~\ref{noise}a.  The scatter of the quadrature amplitudes indicates the total amount of noise added to the signal during preparation, mode conversion, and measurement.  By comparing the total variance, $\mathrm{Var}(X_{1})+\mathrm{Var}(X_{2})$, of the converted signal with that of measured vacuum noise, we infer the mode converter adds $0.92 \pm 0.07$ quanta of noise during conversion with a 90 MHz frequency shift, indicating compatibility with fragile quantum states.   

We have demonstrated a temporal and spectral mode converter whose operating frequency, bandwidth, and low noise properties make it compatible with many circuit QED systems.  As a next step, our mode converter could be integrated with a single cQED system consisting of a superconducting qubit embedded in a high-quality microwave resonator.  This simple network could enable, for example, heralded preparation of multiple phonon Fock states \cite{galland2014, schuster2007}, and serve as a prototype for larger microwave quantum networks.  

\subsection*{Acknowledgments}

This work was supported by the Gordon and Betty Moore Foundation, AFOSR MURI under grant number FA9550-15-1-0015, and the National Science Foundation under grant number 1125844.  A.P.R acknowledges support from the National Science Foundation Graduate Research Fellowship under grant number DGE 1144083.  We thank M.D. Schroer, J. Kerckhoff, and T.A. Palomaki for discussions and advice.  

\clearpage
\bibliography{arxiv-v1-ref}

\onecolumngrid
\newpage
\appendix
\renewcommand\thefigure{\thesection\arabic{figure}}    
\setcounter{figure}{0}    
\renewcommand\thetable{S\arabic{table}}    
\setcounter{table}{0}    
\setcounter{equation}{0}
\setcounter{section}{19}
\section*{Supplementary Information}

\subsection{Measurement network}

\begin{figure}
\begin{minipage}{\linewidth}
\scalebox{0.925}{\includegraphics{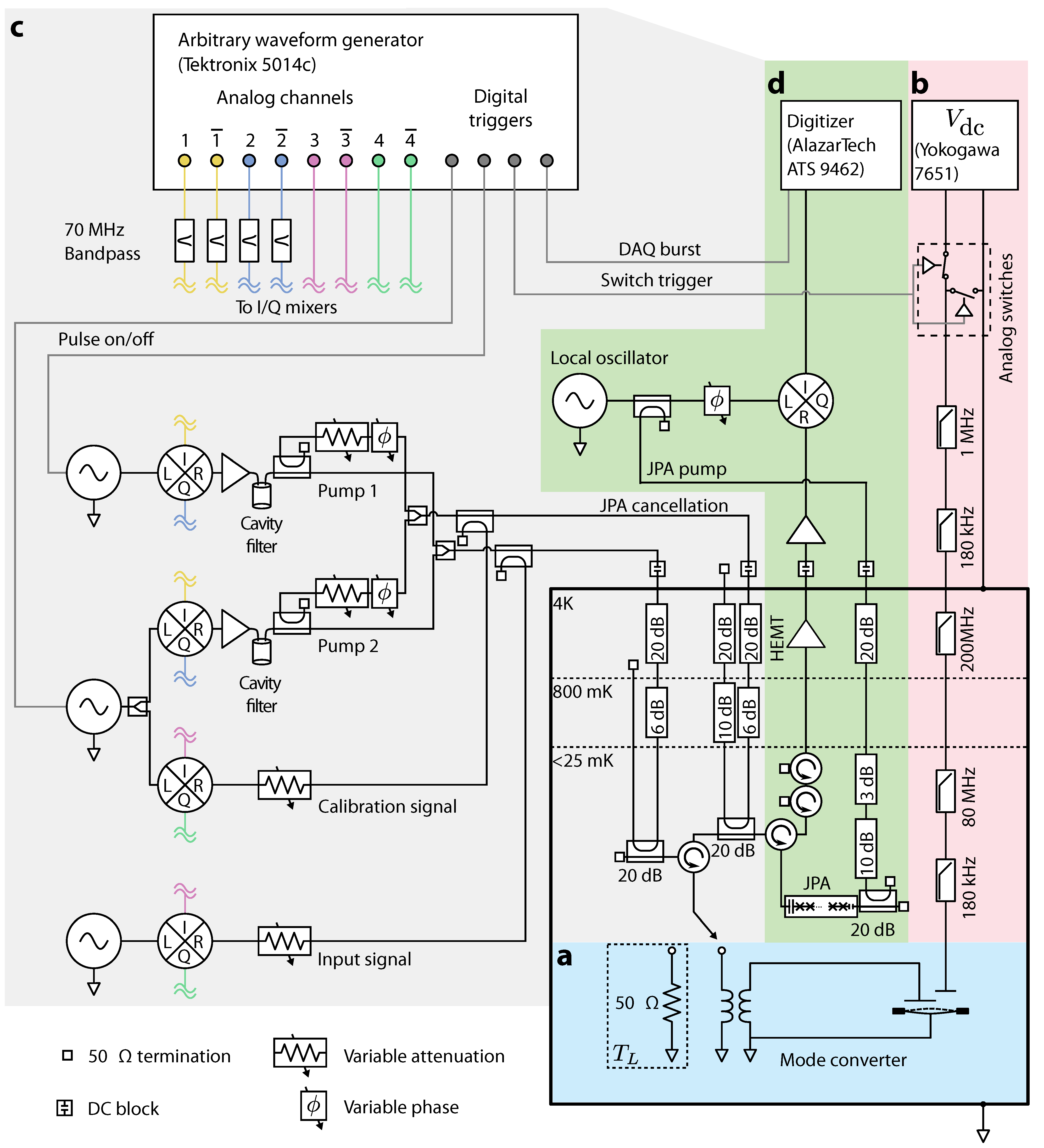}}
\caption{Measurement network. \textbf{a,} Mode converter at $<$25 mK and load resistor at $T_\textrm{L}$.  \textbf{b,} Actuation line. \textbf{c,}  Arbitrary waveform generation for microwave pumps and signals. \textbf{d,} Microwave receiver.}
\label{setup}
\end{minipage}
\end{figure}

Fig. \ref{setup} shows a detailed diagram of the measurement network.  The network consists of four main parts: The mode converter and a load resistor are mounted to the base stage of a dilution refrigerator (subsection \ref{EMC}). A voltage bias on an actuation line controls the center frequency of the mode converter (subsection \ref{dcLine}). Microwave signals control and probe the mode converter (subsection \ref{ssb}). A microwave receiver measures signals after they have interacted with the mode converter (subsection \ref{JPA}).

\subsubsection{Electromechanical circuit and load resistor}
\label{EMC}

The mode converter and a load resistor are mounted to the base stage of a dilution refrigerator and cooled to $<$25 mK.  A low insertion loss ($<$0.30 dB) 2-way switch (Radiall R577433000) connects the microwave receiver to either the mode converter or the load resistor.  The temperature of the load resistor, $T_{\mathrm{L}}$, can be varied without significantly affecting the temperature of the mode converter or fridge, as the load resistor has a weak thermal link to the base of the dilution refrigerator and has an attached heating element.  The load resistor injects a known amount of noise into the microwave receiver, and this noise allows us to calibrate the efficiency and gain of the microwave receiver (see section~\ref{JPA}).

\subsubsection{Actuation line}
\label{dcLine}

The voltage on the actuation line controls the microwave circuit's resonant frequency.  We use a stable voltage source (Yokogawa 7651) to provide a constant $V_{\textrm{dc}}$. Solid state switches (Maxim Integrated 313) change the voltage on the actuation line between $V_{\textrm{dc}}$ and 0~V. The state of the switch network is controlled by a digital output of the AWG (see section~\ref{ssb}).

The actuation line is filtered to ensure voltage fluctuations in $V_{\textrm{dc}}$ do not incoherently drive the mechanical oscillator.  In particular, the mechanical oscillator is sensitive to voltage fluctuations near its resonant frequency of $\omega_{\mathrm{m}}\approx2\pi\times 9$~MHz.  Room temperature filters attenuate fluctuations in $V_{\textrm{dc}}$ by more than 100 dB at frequencies near $\omega_{\mathrm{m}}$, and cryogenic low-pass filters provide additional attenuation near $\omega_{\mathrm{m}}$.  These filters are designed so that the voltage noise seen by the mode converter (at frequencies near $\omega_{\mathrm{m}}$) is ideally given by Johnson noise from a few Ohms at $<$25 mK.  Additionally, the filters provide enough bandwidth to quickly ($<$30~$\mu$s) change the voltage on the actuation line.

\subsubsection{Arbitrary microwave waveform generation}
\label{ssb}

Waveforms with programmable amplitude, phase, and frequency are generated by a Tektronix 5014c arbitrary waveform generator (AWG).  However, the AWG can only produce waveforms with center frequencies up to approximately 300 MHz.  We must upconvert these waveforms so they interact with the mode converter, which has a center frequency of approximately 7 GHz.  Upconversion is accomplished with single-sideband modulation, a process where the original waveform (centered at frequency $\omega_{1}$) modulates a constant microwave tone (at frequency $\omega_{2}$) to produce a final waveform (centered at frequency $\omega_{1}+\omega_{2}$).  Agilent PSG signal generators provide the constant microwave tones, and microwave mixers (Marki Microwave IQ4509-MXP) combine the microwave tones and the waveforms from the AWG.  The relative phase, amplitude, and DC offsets of the AWG waveforms are adjusted to compensate for non-idealities in the mixer.  

Channels 1 and 2 of the AWG produce four waveforms centered at 70 MHz that create the microwave pumps used during our mode conversion protocol.  Bandpass filters (KR Electronics 2657) at the outputs of the AWG attenuate phase and amplitude noise by $>$40~dB for frequencies outside their passband ($\approx$20~MHz centered at 70 MHz).  Although the filters limit the bandwidth of waveforms produced by the AWG, this bandwidth is sufficient for our mode conversion protocol.  Once the waveforms are upconverted using single-sideband modulation, they are amplified (Mini-Circuits ZVA-183V) and additionally filtered \cite{rogers}.  Ideally, this procedure results in a microwave pump at frequency $\omega_{\mathrm{e,1}}-\omega_{\mathrm{m}}$ or frequency $\omega_{\mathrm{e,2}}-\omega_{\mathrm{m}}$ that is $>$170~dB more powerful than at frequency $\omega_{\mathrm{e,1}}$ or $\omega_{\mathrm{e,2}}$.  This large dynamic range is necessary to produce a powerful microwave pump whose phase and amplitude noise do not incoherently drive vibrations of the mechanical oscillator.   When no pump is necessary, the Agilent PSGs are pulsed off so that they emit no microwave tone. 

Channels 3 and 4 of the AWG produce waveforms centered at 30 MHz that create the signals used during our mode conversion protocol.  Once upconverted, the waveforms produced by the AWG are at frequency $\omega_{\mathrm{e},1}$ or $\omega_{\mathrm{e,2}}$.  These signals are neither amplified nor filtered because the signals can be attenuated to reduce phase and amplitude noise.  The signal at frequency $\omega_{\mathrm{e, 1}}$ is manipulated with the mode converter.  The signal at frequency $\omega_{\mathrm{e, 2}}$ is used to monitor the gain of the JPA and additionally used to correct small ($<10$~ps) timing errors that occur between the AWG and the Agilent PSGs during each repetition of our mode conversion protocol.    

\begin{figure}
\begin{minipage}{\linewidth}
\scalebox{0.8}{\includegraphics{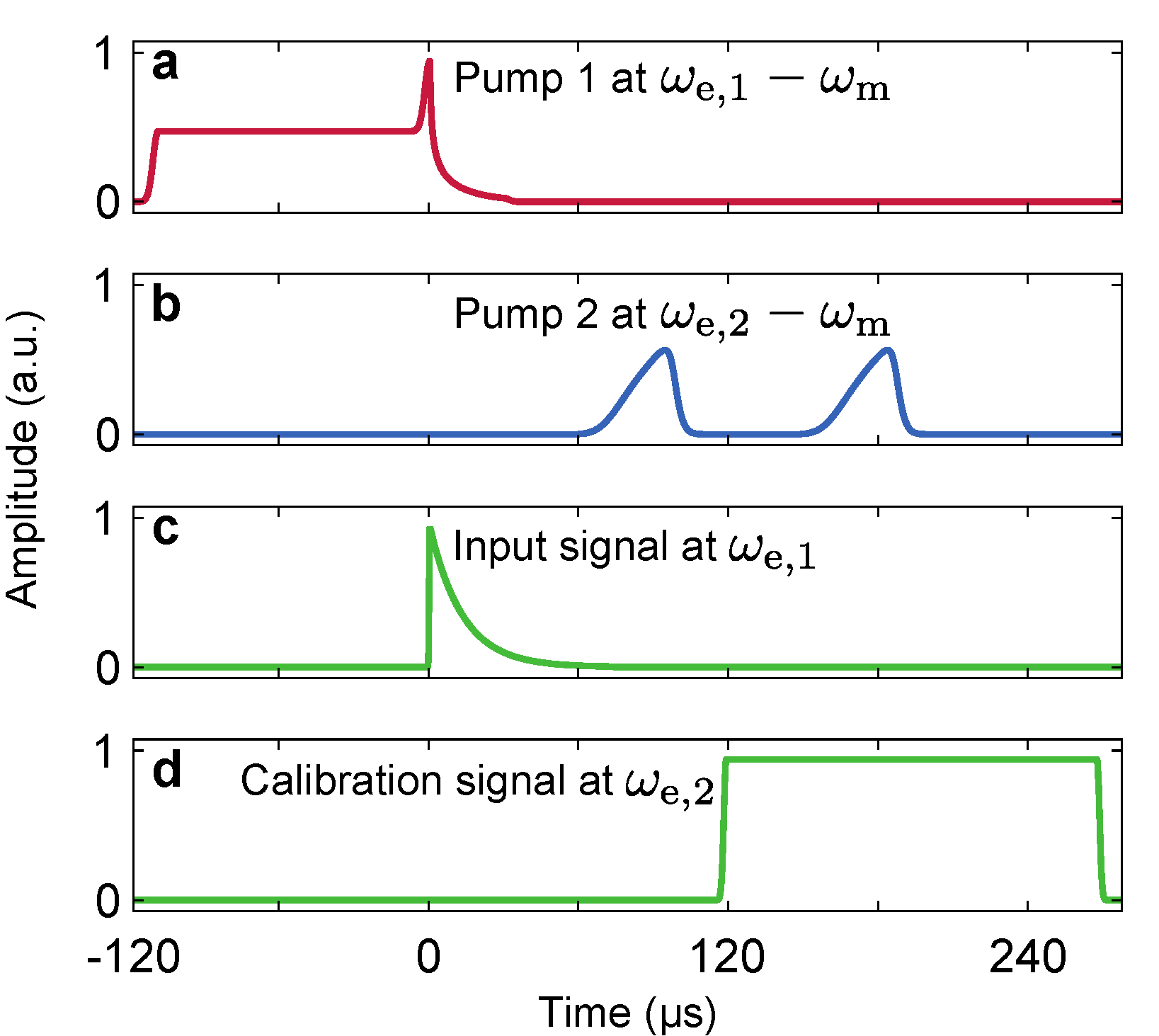}}
\caption{Temporal envelopes of microwave pumps and signals used during temporal and spectral mode conversion.  \textbf{a,} Initially, a constant pump cools the motion of the drumhead.  From $0<t<30~\mu$s, the amplitude of the pump is modulated so that the coupling is given by Eqn.~\ref{decaycatch}.  \textbf{b,} After a 30 $\mu$s storage time ($30<t<60 ~\mu$s), the amplitude of the pump is modulated so the coupling is given by Eqn.~\ref{gaussrelease}.  A second, identical pulse is used in conjunction with the calibration signal to ensure the pump does not saturate the JPA's gain.  \textbf{c,} An exponentially decaying signal at frequency $\omega_{\mathrm{e,1}}$ is injected into the mode converter.  \textbf{d,} A calibration signal at frequency $\omega_{\mathrm{e, 2}}$ is used to monitor the JPA gain and correct for timing errors.   }
\label{pulseSeq}
\end{minipage}
\end{figure}

The temporal envelopes of the microwave pumps and signals produced by the AWG and upconversion are shown in Fig.~\ref{pulseSeq}.

\subsubsection{Microwave receiver}
\label{JPA}

Microwave signals reflected or emitted from the mode converter are measured via a sensitive microwave receiver. The first component of the receiver is a Josephson parametric amplifier (JPA), followed by additional amplifiers, a downconverting mixer, and a digitizer.  A microwave pump detuned 1 MHz below the converter's resonant frequency of $\omega_{\mathrm{e,2}}=2\pi\times7.34$~GHz drives the JPA.  This pump also serves as the phase reference for the downconverting mixer.  We adjust the phase of the pump entering the local oscillator (LO) port of the downconverting mixer so that the signal amplified by the JPA exits via the in-phase channel of the mixer.  This channel is sampled by a high speed digitizer (AlazarTech ATS 9462).

The JPA has a limited dynamic range, and the pumps used with the mode converter carry enough power to saturate the JPA's gain. We use variable attenuators and phase shifters to create cancellation tones that reduce the pump power incident on the JPA by 50 to 60 dB, which is sufficient for proper JPA operation.  We monitor the gain of the JPA (for both gain drift and saturation) with a weak calibration signal (Fig.~\ref{pulseSeq}d).

\subsection{Calibration of microwave receiver}
\label{SII}

The microwave receiver employs a Josephson parametric amplifier (JPA) \cite{manuel2008}.  The JPA amplifies one quadrature of a microwave signal and deamplifies the other, where a quadrature is composed of linear combinations of signals at frequencies above and below the JPA's pump frequency \cite{yurke1989}.  Before amplification or deamplification, the spectral density of each quadrature emitted from the load resistor is given by, in units of quanta (a quanta corresponding to a photon per second per unit bandwidth),
\begin{align}
\label{eff}
\mathcal{S}_{\mathrm{L}}=\frac{1}{2}\mathrm{coth}\left(\frac{\hbar \omega}{2 k_{\mathrm{B}}T_{\mathrm{L}}}\right)
\end{align}
where $T_{\mathrm{L}}$ is the temperature of the load resistor \cite{ manuel2009}.  At temperatures $T_{\mathrm{L}}\ll \hbar\omega/k_{\mathrm{B}}$, $S_{\mathrm{L}}$ approaches 1/2 quantum.  At these low temperatures, the total spectral density of both quadratures is thus 1 quantum, consistent with the expectation that each unit of bandwidth has a minimum spectral density of 1/2 quantum \cite{clerk2010}.  Similarly, at temperatures $T_{\mathrm{L}}\gg \hbar\omega/k_{\mathrm{B}}$, the total power emitted in both quadratures by the resistive load is $2 k_{\mathrm{B}}T_{\mathrm{L}}/\hbar\omega$, the classical Johnson noise formula for two units of bandwidth.

\begin{figure}
\begin{minipage}{\linewidth}
\scalebox{1}{\includegraphics{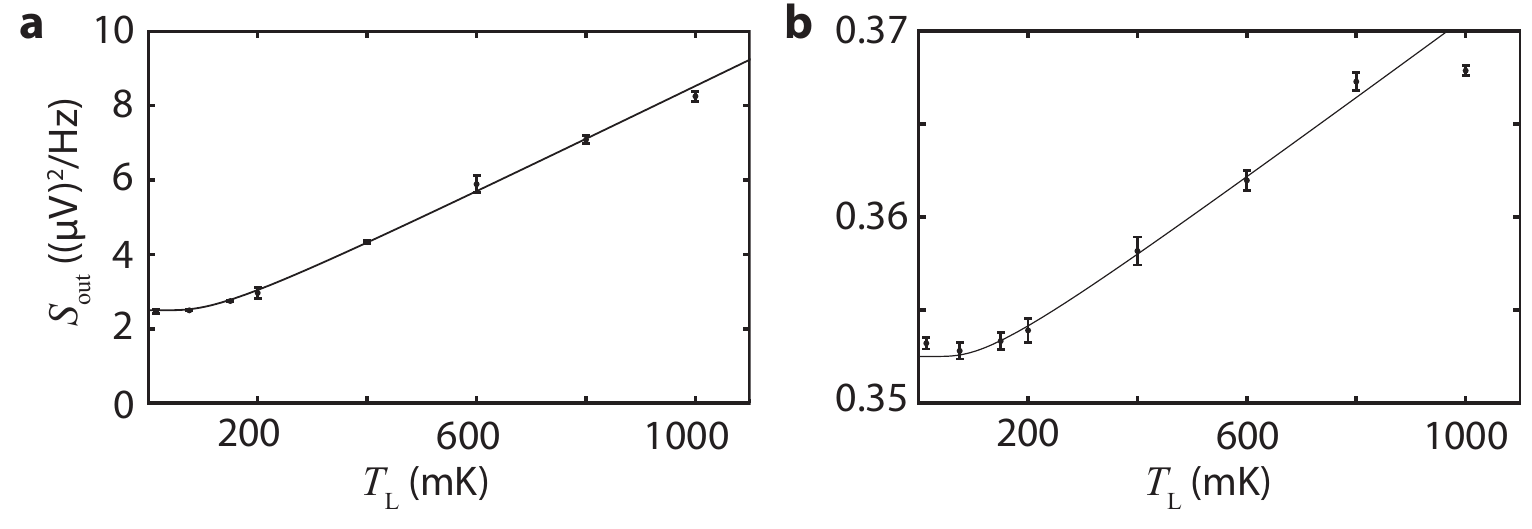}}
\caption{Calibration of microwave receiver. The output spectral density of the microwave receiver varies with the temperature of the load resistor.  {\bf{a}}, A fit to Eqn.~\ref{eff} (solid line) shows the receiver has an efficiency of $\zeta_1=0.49$ and $\mathcal{G}_1=5.0~ (\mu \mathrm{V})^2/(\mathrm{Hz}\cdot\mathrm{quanta})$.  {\bf{b}}, Without the JPA, the efficiency of the receiver drops to $\zeta'_1=0.0105$ and the gain reduces to $\mathcal{G}'_1=0.705~ (\mu \mathrm{V})^2/(\mathrm{Hz}\cdot\mathrm{quanta})$.  }
\label{efficiency}
\end{minipage}
\end{figure}

Once measured by the microwave receiver, we expect to observe a spectral density given by \cite{palomaki2013-2}
\begin{align}
\mathcal{S}_{\mathrm{out}}=\mathcal{G}\left( \frac{  1-\zeta }  {2}+\zeta \mathcal{S}_{\mathrm{L}}          \right)
\end{align}
where $\zeta$ is the efficiency of the receiver and $\mathcal{G}$ is the gain.  The efficiency can range from $\zeta=0$ (no measurement) to $\zeta=1$ (perfect measurement).  Fitting this model to the measured receiver output at $\omega_{\mathrm{e,1}}=2\pi\times 7.08$~GHz yields a gain and efficiency as shown in Fig.~\ref{efficiency}.  

We use the microwave receiver to measure signals reflected or emitted from the mode converter.  These signals are entirely contained in frequencies above the JPA pump frequency.  As such, the measured signal is contained neither in the amplified nor deamplified quadrature, but in both.  The signal in the deamplified quadrature is not measured and this loss of information reduces the efficiency of our microwave receiver by a factor of two, so for a signal near frequency $\omega_{\mathrm{e,1}}$, the efficiency of the receiver is $\zeta/2=0.25$.  The maximum possible efficiency for this mode of operation is $\zeta/2=0.5$.  

During the experiment, we operate the microwave receiver around two different frequencies, $\omega_{\mathrm{e,1}}$ and $\omega_{\mathrm{e,2}}$. The receiver has a frequency dependent performance, so let $\mathcal{G}_{1}$ ($\mathcal{G}_{2})$ be the gain and $\zeta_{1}$ ($\zeta_{2}$) be the efficiency of the receiver at frequency $\omega_{\mathrm{e,1}}$ ($\omega_{\mathrm{e,2}}$). These values are summarized in Table~\ref{gainsJPA}.

\begin{table}
\caption{\label{gainsJPA} Gain and efficiency of the microwave receiver.  }
\begin{ruledtabular}
\begin{tabular}{c  l  l}
{\textbf{Operating Frequency}} & \textbf{Gain} in units of $(\mu\mathrm{V})^2/(\mathrm{Hz}\cdot\mathrm{quanta})$ & {\textbf{Efficiency}} \\
$\omega_{\mathrm{e,1}} = 2 \pi \times $  7.07825 GHz  & $\mathcal{G}_1 =5.0$& $\zeta_1 = 0.49$\\
$\omega_{\mathrm{e,2}} = 2 \pi \times $ 7.34135 GHz & $\mathcal{G}_2 =11$ & $\zeta_2 = 0.60$\\
\end{tabular}
\end{ruledtabular}
\label{params}
\end{table}

To acquire the data in Fig. 3 of the main text, we did not employ the JPA. Without the JPA, the gain and efficiency of the microwave receiver are considerably different, so let $\mathcal{G}'_1$ ($\mathcal{G}'_2$) be the gain and $\zeta'_{1}$ ($\zeta'_{2}$) be the efficiency of the receiver without the JPA at frequency $\omega_{\mathrm{e,1}}$ ($\omega_{\mathrm{e,2}}$); these values are summarized in Table~\ref{gainsNoJPA}. To calculate the photon number conversion efficiency of the mode converter quoted in the main text, we compared the energy in the unconverted signal at $\omega_{\textrm{e},1}$ and the energy in the converted signal at $\omega_{\textrm{e,2}}$. We calculate a conversion efficiency given by 
\begin{align}
\frac{E_{\mathrm{2}}}{E_{\mathrm{1}}}\times \frac{\mathcal{G}'_{\mathrm{1}}\zeta'_{\mathrm{1}}}{\mathcal{G}'_{\mathrm{2}}\zeta'_{\mathrm{2}}}=\frac{E_{\mathrm{2}}}{E_{\mathrm{1}}} \times 0.60 , \nonumber
\end{align}
where $E_{\mathrm{2}}$~($E_{\mathrm{1}}$) is the energy of the converted~(unconverted) signal measured at the output of the receiver. 

\begin{table}
\caption{\label{gainsNoJPA}Gain and efficiency of the microwave receiver without the JPA.}
\begin{ruledtabular}
\begin{tabular}{c  l  l}
{\textbf{Operating Frequency}} & \textbf{Gain} in units of $(\mu\mathrm{V})^2/(\mathrm{Hz}\cdot\mathrm{quanta})$ & {\textbf{Efficiency}} \\
$\omega_{\mathrm{e,1}} = 2 \pi \times $  7.07825 GHz  & $\mathcal{G}'_1 =0.705$ & $\zeta'_1 = 0.0105$\\
$\omega_{\mathrm{e,2}} = 2 \pi \times $ 7.34135 GHz & $\mathcal{G}'_2 =0.98$ & $\zeta'_2 = 0.0125$\\
\end{tabular}
\end{ruledtabular}
\label{params}
\end{table}

\subsubsection{Quadrature amplitude extraction}
\label{quadrature}

We can express a continuous wave signal in units of quanta using a Fourier transform of the measured voltage waveform.  For such a signal, we define quadrature amplitudes at a frequency $\omega$: 
\begin{align}
\label{eq:quadamp1}
X_1&=\sqrt{\frac{2}{\zeta \mathcal{G}}} \cdot \sqrt{\frac{\Delta t}{N}} \cdot \sum\limits_{k=1}^{N}  y(k) \cdot \mathrm{cos} (\omega \cdot k \cdot \Delta t) \nonumber \\
X_2&=\sqrt{\frac{2}{\zeta \mathcal{G}}}\cdot \sqrt{\frac{\Delta t}{N}}\cdot\sum\limits_{k=1}^{N}  y(k)\cdot\mathrm{sin}(\omega\cdot k\cdot \Delta t)
\end{align}
where $y(k)$ is the measured voltage waveform of length $N$, $1/\Delta t$ is the sample rate, and $\zeta$ is the measurement efficiency (see subsection \ref{JPA}).  The factor $\sqrt{2/(\zeta\mathcal{G})}$ converts the voltage waveform into units of $\sqrt{ \mathrm{Hz}\cdot\mathrm{quanta}}$ at the input of the receiver (the factor of two remains because $\mathcal{G}$ was calculated using a single-sided spectral density).  The quadrature amplitudes $X_1$ and $X_2$ are proportional to the real and imaginary parts of the complex Fourier amplitude, and can be considered to represent the magnitude and phase of the signal: $X_1^2+X_2^2$ yields an estimate of the number of quanta contained in the signal, and $\tan^{-1}(X_2/X_1)$ the phase of the signal.  

Eqns. \ref{eq:quadamp1} are appropriate when the signal is continuous and at a known frequency.  However, when a signal has a particular temporal envelope, this definition no longer provides an accurate description of the signal.  To compensate for the finite bandwidth of the signal, we define quadrature amplitudes using a Fourier transform with a window function that matches the temporal envelope of the signal. This window function prevents the quadrature amplitudes from being affected by noise that occurs either before or after the signal. Including the window function in Eqns. \ref{eq:quadamp1} gives
\begin{align}
\label{eq:quadamp2}
X_1&=\sqrt{\frac{2}{\zeta \mathcal{G}}}\cdot \sqrt{\frac{\Delta t}{    C   }}\cdot\sum\limits_{k=1}^{N} f(k)\cdot  y(k)\cdot\mathrm{cos}(\omega\cdot k\cdot \Delta t) \nonumber \\
X_2&=\sqrt{\frac{2}{\zeta \mathcal{G}}}\cdot \sqrt{\frac{\Delta t}{C}}\cdot\sum\limits_{k=1}^{N} f(k)\cdot  y(k)\cdot\mathrm{sin}(\omega\cdot k\cdot \Delta t)
\end{align}
where $f(k)$ is the temporal envelope of the signal and $C=\sum\limits_{k=1}^{N}  \left|  f(k)  \right|^2$ is a factor that compensates for the noise equivalent bandwidth of $f(k)$.

Using the window function $f(k)$ with a Fourier transform is equivalent to using least-squares parameter estimation, where the parameters to be estimated are the amplitudes of $f(k)\cdot \mathrm{cos}(w\cdot k\cdot \Delta t)$ and $f(k)\cdot \mathrm{sin}(w\cdot k\cdot \Delta t)$.  This method of parameter estimation is appropriate when the noise in the voltage waveform is Gaussian, as is the case for our measurements \cite{gibbs}.   

Using knowledge of the efficiency and gain of our receiver, and knowledge of the temporal envelope and center frequency of the converted signal, we use Eqns. \ref{eq:quadamp2} to calculate the quadrature amplitudes $X_1$ and $X_2$ that are plotted in Fig.~4 of the main text.  

\subsection{Parameters of electromechanical circuit}
\label{tsmcparams}

\begin{table}
\caption{\label{parameters}Parameters of electromechanical circuit. }
\begin{ruledtabular}
\begin{tabular}{c  l  l }
{\bf{Symbol}}	&	{\bf{Description}}	&	{\bf{Value and units}} \\
$\omega_{\mathrm{e}}$	&	Circuit resonant frequency (at $V_{\mathrm{dc}}=0$)	&	$2\pi\times$7.34 GHz	\\
$\kappa$	&	Circuit decay rate	&	$2\pi\times$2.5 MHz	\\
$\kappa_{\mathrm{ext}}$	 &	Circuit decay rate into transmission line	&	0.92$\times\kappa$		\\
$\omega_{\mathrm{m}}$ & Mechanical resonant frequency (at $V_{\mathrm{dc}}=0$)&  $2\pi\times$9.56 MHz \\  
$\kappa_\mathrm{m}$ & Mechanical decay rate& $2\pi\times$25 Hz  \\
$n_{\mathrm{m}}$& Average occupancy of the mechanical oscillator	& 36	\\
$n_{\mathrm{m}}\kappa_\mathrm{m}$	& Mechanical decoherence rate	& $2\pi\times$900 Hz	\\
$g_0 = Gx_{\mathrm{zp}}$	&	Electromechanical coupling	&	$2\pi\times$270 Hz			
\end{tabular}
\end{ruledtabular}
\label{params}
\end{table}

Parameters of the microwave circuit, detailed in Table \ref{params}, are determined by injecting a microwave signal and sweeping it over a large range of frequencies.  Fitting the reflected microwave signal allows us to extract the resonant frequency of the microwave circuit ($\omega_{\mathrm{e}}$) and the external coupling rate ($\kappa_{\mathrm{ext}}$), and total energy loss rate ($\kappa$).  

Properties of the mechanical oscillator are also determined using the microwave response of the device.  A microwave pump is injected into the mode converter, and motion of the aluminum drumhead modulates the pump and produces modulation sidebands on the reflected pump \cite{schliesser2008-2}.  The frequency and spectral width of the sidebands indicate the vibrational frequency of the drumhead ($\omega_{\mathrm{m}}$) and (at low microwave power) its intrinsic damping rate ($\kappa_\mathrm{m}$).  The magnitude of the sidebands indicates the coupling strength between the microwave circuit and the aluminum drumhead.  The amount of power in the sideband ($P_{\mathrm{m}}$) at a frequency $\omega_{\mathrm{m}}$ above the injected microwave pump is proportional to the amount of power in the injected microwave pump ($P_{\mathrm{c}}$) \cite{rocheleau2010}: 

\begin{align}
\frac{P_{\mathrm{m}}}{P_{\mathrm{c}}}=\frac{g_0^2}{2}\frac{\left<  x^2 \right>}{ x_{\mathrm{zp}}^2}  \frac{\kappa_{\mathrm{e}}^2}{\Delta^2+\left(  \kappa_{\mathrm{e}}-\kappa/2  \right)^2} \frac{1}{\left( \Delta+\omega_{\mathrm{m}}  \right)^2  + \left(  \kappa/2 \right)^2}
\end{align}
where $\Delta$ is the difference between the pump frequency and the microwave circuit's resonant frequency, $x_{\mathrm{zp}}$ is the zero-point fluctuations of the drumhead, and $\left<  x^2 \right>$ describes the motion of the drumhead.  For a drumhead in equilibrium with the thermal environment provided by the dilution fridge, $\left< x^2 \right>=2 x_{\mathrm{zp}}^2  n_{\mathrm{m}}$, where $n_{\mathrm{m}} \approx  k_{\mathrm{B}}T /\hbar\omega_{\mathrm{m}} $ is the average phonon occupation of the drumhead for dilution fridge temperature $T$.  

\begin{figure}
\begin{minipage}{\linewidth}
\scalebox{1}{\includegraphics{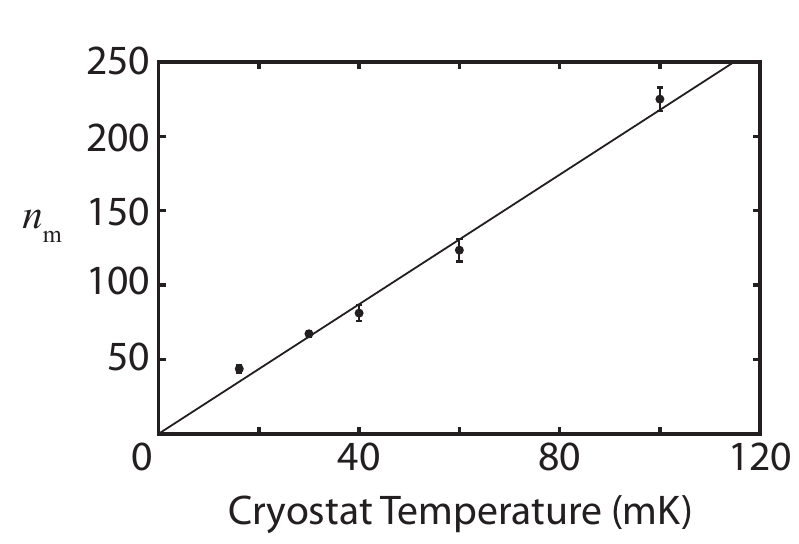}}
\caption{Electromechanical coupling and thermalization of the aluminum drumhead.  Black line shows expected phonon occupancy, $n_{\mathrm{m}}$, given equipartition and an electromechanical coupling rate of $g_0=2\pi\times 270$ Hz. }
\label{gcal}
\end{minipage}
\end{figure}

We vary the fridge temperature and monitor the ratio $P_{\mathrm{m}}/P_{\mathrm{c}}$, expressed as an effective phonon occupation, $n_{\mathrm{m}}$.  As shown in Fig. \ref{gcal}, at high temperatures (50-100 mK) the drumhead is in thermal equilibrium with the fridge as evidenced by the linear relationship between fridge temperature and $n_{\mathrm{m}}$.  This relationship allows us to extract an electromechanical coupling of  $g_0=2\pi\times 270$ Hz.  At the fridge base temperature of approximately 13 mK, the phonon occupation number saturates to $n_{\mathrm{m}}\approx 36$.    

\subsubsection{Expected frequency tuning of microwave circuit}

\begin{figure}
\begin{minipage}{\linewidth}
\scalebox{.4}{\includegraphics{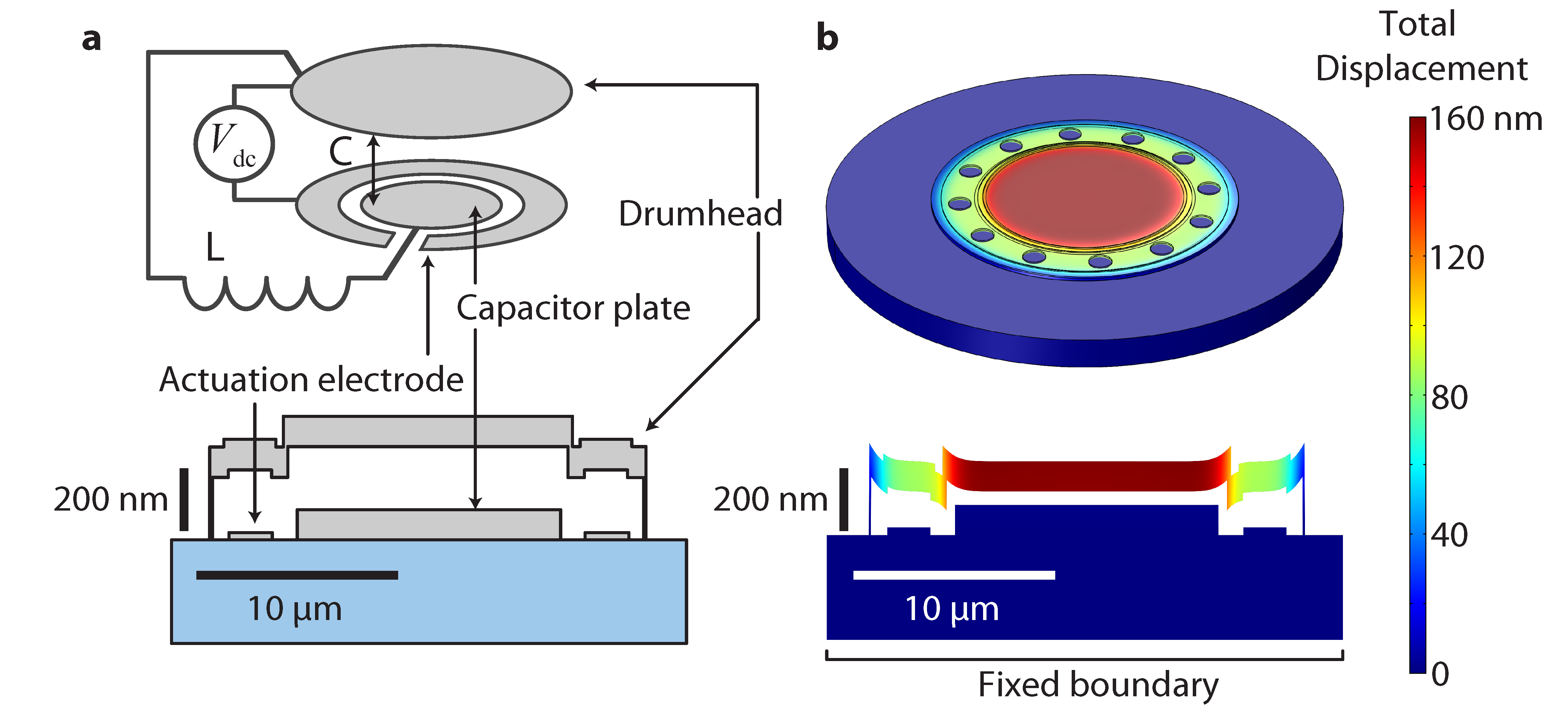}}
\caption{Mode converter geometry.  {\bf{a}},   Diagram of the mode converter, and cross section of the parallel plate capacitor, which is formed by depositing aluminum films (gray) on a sapphire substrate (light blue).  At room temperature, there is approximately 200 nm of separation between the two metal films.  {\bf{b}}, Finite element simulation of device geometry after cooling from room temperature to $<$25 mK.  }
\label{comsolv1}
\end{minipage}
\end{figure}

The approximate geometry of the aluminum drumhead is outlined in Fig.~\ref{comsolv1}a.  With this geometry, we use a finite element simulation to predict the effect of thermal contraction between the aluminum and the sapphire substrate.  In simulation, the total thermal contraction is adjusted so that the central portion of the drumhead deflects by $\sim$160 nm, leaving a $\sim$40 nm separation between the plates of the microwave capacitor, a value set by the measured $g_0 = 2\pi \times 270$ Hz.  The total thermal contraction needed to create this amount of deflection is seven parts per thousand, a value approximately consistent with that of bulk aluminum.  The predicted effect of thermal contraction on device geometry is shown in Fig.~\ref{comsolv1}b.   

The predicted thermal contraction of Fig.~\ref{comsolv1}b shows approximately 120 nm separates the drumhead from the actuation electrode used to deflect the drumhead.  Using this thermally-altered geometry, we again use a finite element simulation to predict the effect of an applied voltage on drumhead deflection and resonant frequency.  As the drumhead deflects with increasing voltage, it alters the capacitance in the microwave circuit and thus its resonant frequency.  This effect is modeled by assuming a parallel plate capacitance ($C_{\mathrm{m}}$) that is modulated by changes in the position of the drumhead, and an additional parasitic capacitance ($C_{\mathrm{p}}$) provided by, for example, the coils of the inductor.  With no applied voltage, $C_{\mathrm{m}}/\left(  C_{\mathrm{p}} + C_{\mathrm{m}}   \right)=0.65$, estimated from electromagnetic simulations of the microwave circuit.  Using this information, we predict the frequency tuning of the device shown as dashed lines in Fig. 1a and Fig. 1b of the main text.

In order to correctly predict the frequency tuning of the device, we must include the effects of the Casimir force.  We include an attractive pressure of approximately $0.7\times P_{\mathrm{cas}}$ in a finite element simulation of the membrane in the presence of a DC voltage, where $P_{\mathrm{cas}}=-\hbar c \pi^2/240 d^4$ is the Casimir pressure for two parallel, perfectly conducting plates separated by a distance $d$ of vacuum.  (For imperfectly conducting plates, a reduction in the Casimir force for small separations $d\lesssim\lambda_{\mathrm{p}}$ is expected, where $\lambda_{\mathrm{p}}$ is the plasma wavelength of the conducting plates \cite{intravaia2005}.)  With this additional pressure, we predict the frequency tuning of the device shown as solid lines in Fig. 1a and Fig. 1b of the main text.  

\begin{figure}
\begin{minipage}{\linewidth}
\scalebox{.6}{\includegraphics{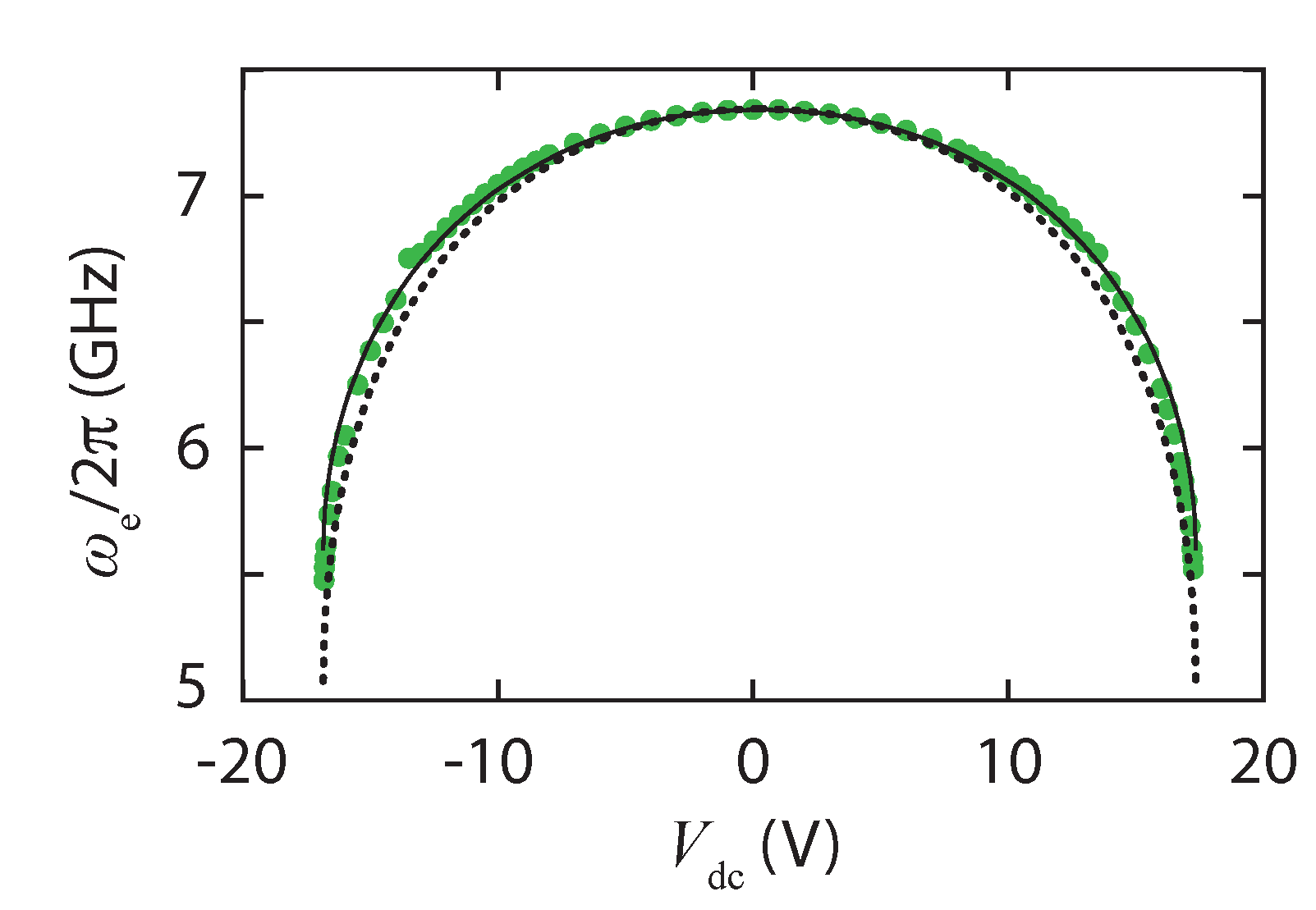}}
\caption{Measured microwave resonant frequency ($\omega_{\mathrm{e}}$) as a function of $V_{\mathrm{dc}}$.  The solid line shows the predicted frequency tuning as calculated with a finite element simulation that includes a Casimir force.  The dotted line shows the predicted frequency tuning as calculated with a simulation that includes a 0.75 V potential difference between the drumhead and the capacitor plate (but no Casimir force).  }
\label{patchpot}
\end{minipage}
\end{figure}

Although inclusion of a Casimir force explains our data reasonably well, attractive forces other than the Casimir force could potentially explain the observed frequency tuning of the device.  One such attractive force is provided by patch potentials \cite{speake2003}.  A finite element simulation that includes a 0.75 V potential difference between the drumhead and the capacitor plate predicts a frequency tuning shown by the dotted line in Fig.~\ref{patchpot}.  This prediction does not explain our data as well as inclusion of an attractive force that scales like the Casimir force.  However, the observed frequency tuning of the device could be explained by a combination of both the Casimir force and forces from patch potentials.  

\subsection{Shape of $\Gamma(t)$}

Controlling the temporal envelope of microwave signals using our converter requires optimally shaping $\Gamma(t)$. This section provides an overview of how we choose a suitable $\Gamma(t)$ and use our mode converter to capture and release microwave signals with arbitrary temporal envelopes.

For a cavity optomechanical system in the presence of a strong, monochromatic pump, the equations of motion for the position of the mechanical oscillator (described by $x=x_{\mathrm{zp}}(c+c^*)$) and the fluctuations of the resonator field about the pump (described by complex mode amplitude $a$) are \cite{kippenberg2013}
\begin{align}
\label{lineareq1}
\dot{a}(t)&=(i\Delta-\kappa/2)a(t)-i g (c(t)+c^* (t))+\sqrt{\kappa_{\mathrm{ext}}}a_{\mathrm{in}}(t) \nonumber \\
\dot{c}(t)&=(-i\omega_{\mathrm{m}}-\kappa_{\mathrm{m}}/2)c(t)-i\left(g a^*(t)+g^* a(t)\right) \nonumber \\
a_{\mathrm{out}}(t)&=\sqrt{\kappa_{\mathrm{ext}}}a(t)-a_{\mathrm{in}}(t)
\end{align} 
where $\Delta$ is the difference between the pump frequency and the microwave circuit's resonant frequency, and other parameters are as given in Table~\ref{params} and discussed in section~\ref{tsmcparams}.  The amplitude $a_{\mathrm{in}}(t)$ describes the signal incident on the microwave circuit, and the coupling $g=g_0 \alpha$, where $|\alpha|^2$ is the number of photons induced in the microwave circuit by the pump.  Crucially, the coupling $g$ can be modulated by changing the amplitude of the pump.  
If we fix the detuning to $\Delta=-\omega_{\mathrm{m}}$, move into a frame rotating at the microwave circuit's resonant frequency, and make a resolved-sideband approximation ($4\omega_{\mathrm{m}}\gg\kappa$) and a weak-coupling approximation ($g\ll\kappa$), the equations of motion reduce to \cite{zhang2003, harlow2013}
\begin{align}
\label{mechcatch}
\dot{c}(t)&=-\frac{1}{2}\Gamma(t)  c(t)  -e^{-i\psi}\sqrt{\eta\Gamma(t)}a_{\mathrm{in}}(t) \nonumber\\
a_{\mathrm{out}}(t)&=e^{i\psi}\sqrt{\eta\Gamma(t)}c(t)+(2\eta-1)a_{\mathrm{in}}(t)
\end{align}
where $\Gamma(t)=4|g(t)|^2/\kappa\gg\kappa_{\mathrm{m}}$, $\eta=\kappa_{\mathrm{ext}}/\kappa$, and $\psi=\mathrm{Arg}(-i g)$.  

Eqns.~\ref{mechcatch} succinctly describe how an incoming signal, $a_{\mathrm{in}}(t)$, couples to the motion of the drumhead used in our mode converter.  For $\eta\approx1$, optimally capturing an incoming signal as a vibration of the drumhead is equivalent to minimizing the reflected signal, $a_{\mathrm{out}}(t)$.   Minimizing the energy in the reflected signal with $\eta=1$ requires a coupling $\Gamma(t)$ of \cite{harlow2013} 
\begin{align}
\label{optimalcatch}
\Gamma(t) = { e^{\kappa_\mathrm{m} t } |a_\mathrm{in}(t)|^2 \over \frac{a_\mathrm{in}(0)^2}{\Gamma(0)} +  \int_0^t e^{\kappa_\mathrm{m} t'} |a_\mathrm{in}(t')|^2 dt' }
\end{align}
where $\Gamma(0)$ is the coupling at $t=0$.  This coupling can be found by using the Euler-Lagrange equation to find a stationary point of the reflected energy, $\int_0^{T} |a_{\mathrm{out}}(t)|^2dt$, with $a_{\mathrm{out}}(t)$ as given by Eqns.~\ref{mechcatch}.  

In Fig. 3 of the main text, we capture an incoming signal with an exponentially decaying temporal envelope (i.e., $|a_{\mathrm{in}}|\propto \Theta(t) e^{-\gamma t/2}$, where $\Theta(t)$ is the Heaviside step function) as a vibration of the drumhead.  Optimally capturing this signal requires a $\Gamma(t)$ as prescribed by Eqn.~\ref{optimalcatch}: 
\begin{align}
\label{decaycatch}
\Gamma(t)  =   \Theta(t) \frac{\gamma e^{-\gamma t}} {1 - e^{-\gamma t} + \gamma/ \Gamma(0)}
\end{align}
where $\Gamma(0)$ is an input; $\Gamma(0)\gg\gamma$ yields a high capture efficiency.  We use $\gamma=2\pi\times 24$~kHz and $\Gamma(0)\approx 2\pi\times 500$~kHz.  Ultimately, $\Gamma(t)$ is limited by the microwave circuit's bandwidth to $\Gamma(t)<\kappa/2$.  

Once the incoming signal is captured, it is temporarily stored and then released.  During storage, $\Gamma(t)=0$.  To release a signal with a Gaussian temporal envelope centered at time $t_{\mathrm{o}}$ (i.e., $|a_{\mathrm{out}}(t)|\propto e^{-\gamma^2(t-t_{\mathrm{o}})^2}$), we use Eqn.~\ref{optimalcatch} to determine the optimal choice of $\Gamma(t)$: 
\begin{align}
\label{gaussrelease}
\Gamma(t) = \frac{  \gamma e^{-\gamma^2(t-t_{\mathrm{o}})^2}       }{1-\mathrm{erf}\left(    \gamma(t-t_{\mathrm{o}})   \right)+\delta      }
\end{align}
where  erf($t$) is the error function, $\gamma=2\pi\times24$~kHz and $\delta\approx\gamma/\Gamma(0)e^{-\gamma^2t_{\mathrm{o}}^2}$ \cite{harlow2013}.  

\end{document}